# A Parallel IFC Normalization Algorithm for Incremental Storage and Version Control

Liu H.[a,b,*], Gao G.[b,c,*], Gu M.[b,c]

[a] Digital Horizon Technology Co., Ltd., China. [b] Tsinghua University, China. [c] Beijing National Research Center for Information Science and Technology (BNRist), China.

liuhan@digital-h.cn; gaoge@tsinghua.edu.cn

**Abstract.** Industry Foundation Classes (IFC) files are commonly used for data exchange of Building Information Models (BIMs). Due to the equivalent transformations in the graph structure of IFC data, it is a challenge to perform version comparison and incremental storage on IFC files. In this paper, an IFC normalization method is proposed, which can reduce the influence of the equivalent transformations, so that the normalized IFC file can be directly used in Git-like tools for version comparison and incremental storage. The algorithm is also designed for getting stable results when running on multi-threads. Experiments show the efficiency of the algorithm and its potential in Common Data Environment (CDE) applications.

## 1. Introduction

Building Information Model (BIM) is the digital representation of building projects and built assets in the construction industry. The ISO-19650 standard recommends BIM-based information management in the life cycle of a built asset, which includes the design, construction, and operation phases. During the life cycle, the BIM data is created and edited continuously by multiple parties, and information is accumulated in multiple stages of the whole construction process. The Common Data Environment (CDE) is a type of information system for sharing and coordinating BIM data among multiple parties, for realizing the BIM-based information management. Version control is an important function of the CDE, in order that the BIM data can be iteratively modified and synchronized between multiple parties, and the historical changes of data can be tracked.

The Industry Foundation Classes (IFC) data model is commonly used for BIM data exchange. The IFC schema defines the classes of data nodes for representing various facets of information such as product types, geometry, properties and relationships. The data nodes form a directed acyclic graph (DAG) structure. DAGs are without cycled references, and the data of every node is determined by a subgraph rooted with the node. The graph is serialized into ASCII text rows for data storage in hard drives and data transmission through the internet. During the whole process of a construction project, the BIM data can reach tens of gigabytes or more. Such a large amount of data needs to be iteratively modified and synchronized between multiple parties. If the complete IFC files are stored and exchanged for each version, the disk space occupation and the time usage for transmission will be large. Therefore, it is of great value to implement version comparison and incremental storage for BIM data.

Currently, in the information technology industry, various tools like Git and SVN are commonly used in software programming projects for version control, incremental storage and code synchronization. There are also mature online platforms developed based on such version control tools, which are widely used for the collaboration of multiple software developers. However, although the IFC file is also in a text format, such Git-like version control tools can not be directly used on IFC files. One important reason is that the IFC schema is defined in a graph structure, but the current version control tools focus on the serialized linear file contents.

Due to the equivalent transformations in the IFC file, an identical graph structure may be serialized into totally different text files, which brings problems for Git-like tools to compare the IFC data versions.

Several related studies have tried to develop new version comparison and incremental storage tools specifically for the graph structure of IFC data. However, on the one hand, the space and time efficiency of graph-based incremental storage is still a challenge; on the other hand, there is still a lot of work to develop new user-friendly graph-based version control systems that is comparable with the mature Git-based platforms like GitLab.

To address this issue, the idea in this paper is to propose a normalization algorithm for the IFC data. Data normalization is a preprocessing step to transform the data into a standardized form so that the normalized data can better fit the subsequent calculation or analysis. The proposed IFC normalization algorithm tries to uniformly serialize the graph structures into linear text files, so that the text-based version control tools can identify the true changes in the normalized IFC files, which makes it possible to use online version control services for recording the version history and synchronizing data with other users.

## 2. Related Work

### 2.1 Common Data Environment for Synchronization and Archiving of BIMs

According to the ISO-19650 standard, CDE is the center of the information management workflow for buildings and infrastructures. The workflow is about the accumulation of data through the phases in the life cycle, the exchange of data between heterogeneous systems, and coordination between multiple parties. The CDE system acts as a single source of BIM data for all parties, and the ability to keep track of the information transactions is defined as an important feature of the CDE. For CDE systems, the efficiency of data synchronization, the traceability of history archives, and the cost of storage are several major challenges (Tao, et al., 2021; Jaskula, et al., 2022).

Several CDE implementations based on different technology stacks have been proposed. The OpenCDE API project (buildingSMART, 2019) tries to provide a standard specification for the behavior of CDE implementations, including authentication, data transferring and synchronization. According to the OpenCDE API, centralized and distributed CDEs can be implemented. A centralized CDE can be based on a file server to store the versions of models, documents and issues, then the end-users can access and update the central data through HTTP API requests (Preidel, et al., 2017; Patacas, et al., 2020). One possible implementation of a distributed CDE is based on the Solid ecosystem (Werbrouck, et al., 2019; Senthilvel, et al., 2020), through which the users can store the data in their own servers, and the remote users can access the linked data sources with distributed authentication. Another idea for distributed CDE is based on the IPFS (InterPlanetary File System), which incrementally archive and share the data in peer-to-peer network, and records hash indices of data versions into blockchain transactions (Tao, et al., 2021). More information of related studies on CDE applications may refer to the literature review (Jaskula, et al., 2022).

### 2.2 Comparison Algorithms for BIM Versions

The graph structure of the IFC data needs to be serialized into a linear file format for storage and transmission. The most used file format for the IFC schema is based on the ISO 10303-21 standard. The data nodes are serialized into text rows, and a node can be referred to by other nodes with the integer row ID. When serializing the graph structures, the following equivalent transformations may occur.

- Storage order change: the data rows are shuffled in the IFC file, or the values in an unordered set are shuffled.

- Row ID replaces: the integer ID of a row and all its references in the file are replaced with another integer value.

- Redundant nodes: identical nodes or identical node trees repeat several times in the file.

Due to the equivalent transformations, an identical graph structure may be serialized into totally different text files, which makes the text-based comparison tools fail to identify the true changes between versions of IFC files. Aiming at the challenges, several studies have proposed graph-based comparison algorithms for BIM versions, in order to realize more efficient storage and transmission of BIM data.

One idea is to extract information from the IFC file and store it in some other data structures for version comparison. The ObjectVCS method (Firmenich, et al., 2005; Nour, et al., 2006) extracts necessary information from the IFC file and reorganizes the data of each object into an XML file named with the GUID of the object. When the BIM data is modified, the extracted data files are updated so that the modification can be tracked. A similar strategy is also used in the Speckle system (Poinet, et al., 2020), which implements the version control for BIM data based on the extracted JSON data structure. Comparing the extracted information is a pragmatic way to implement version control for BIM data, but it may be problematic to restore the whole original IFC file from the extracted data files, which limits the use cases of such methods.

Another idea is to compare the BIM data based on the graph structure of the IFC schema. One graph-based version control algorithm (Esser, et al., 2022) compares the IFC graph data by matching the maximum common subgraphs on two versions of the graph data, so that the modifications over the graph structure can be identified. Based on the comparison, graph patches can be generated to synchronize the graph data changes between multiple parties. Another graph-based BIM incremental storage system (Li, et al., 2022) is based on the IFCdiff algorithm (Shi, et al., 2018) for content-based comparison of IFC data nodes. IFCdiff calculates the hash string of each data node from the content of the whole subgraph rooted with the node, and the result hash string is invariant under the equivalent transformations listed above. As a result, the set of all hash strings from the IFCdiff algorithm can be used as a reliable identifier of an IFC file, and the versions can be compared by subtracting two sets of hash strings.

The graph-based version control for IFC is able to identify the true changes in the BIM data. However, in order to compare a new version of graph data with the previous versions, the current methods rely on persistent databases (such as Neo4j and MongoDB) to store the graph structure, node indices and hash results. In a scenario with multiple users, everyone needs to synchronize the database first before comparing a new version of BIM data, which may limit the efficiency and flexibility in distributed applications.

Although faced with challenges, it is still an attractive idea to record and share IFC file versions in Git-like tools. The "Native IFC" specification (Bruno, 2022) is proposed as an initiative for BIM software tools to realize IFC version control on Git-like tools. The Native IFC specification requires that: (1) the numeric row ID of a data entry must keep the same in each edition; (2) the attribute changes must be written in-place; and (3) the IDs of deleted entities must not be reused in later versions. This is a strict but still possible specification for software tools directly editing the IFC data structure. However, for most other BIM tools that need to convert data between another data schema in exporting and importing IFC files, the Native IFC is much more difficult to be implemented.

In this paper, the idea is that although the exported IFC file may be different in text, an IFC normalization algorithm is able to merge the redundant nodes and to reorganize the rows and numerical IDs into a stable equivalent form. As a result, the true modifications can be identified by Git-like tools. The proposed normalization algorithm can be performed independently on any single IFC file without a persistent database to store the historical hash results. In a scenario with multiple users, the normalization can be performed on every single device, which makes it possible to realize BIM-based collaboration on mature Git-like version control platforms.

## 3. The Parallel IFC Normalization Algorithm

In this section, the parallel IFC normalization algorithm is introduced in detail. Inspired by the IFCdiff algorithm, the proposed algorithm is also based on the hash calculation of layered data nodes. The basic idea is to return integer hash codes rather than hash strings for each data node. By assigning content-based integer hash codes as row IDs, the data rows can be sorted and compared in version control systems for text files.

The algorithm for calculating the hash strings and integer hash codes for IFC data nodes in the graph structure is introduced in section 3.1. The "prefix spaces" for resolving hash collision of short integer hash codes in parallel is introduced in section 3.2. The steps of the parallel IFC normalization algorithm are presented in section 3.3. Some special strategies for better dealing with the data in the IFC schema are introduced in section 3.4.

### 3.1 Calculating Hash Strings and Hash Codes for Layered Data Nodes

In the proposed method, both long hash strings and short integer hash codes are used for the IFC data nodes. The long hash strings are reliable identifiers of contents, and the short integer hash codes are used for assigning locations in linear file storage.

The calculation of hash strings is based on the IFCdiff algorithm, which calculates the hash string of a node according to the content of the subgraph rooted with the node. The nodes in a DAG structure can be organized in layers according to the reference edges between the nodes. The bottom layer is composed of all leaf nodes. Each leaf node has only pure literal values but does not refer to other nodes. Marking the bottom layer as layer 0, the other nodes are divided into layers by the following rule: if the maximum layer of the nodes that one node refers to is layer $i$, then the layer of the current node is $i + 1$.

An example of layered hash string calculation is shown in Figure 1. The hash calculation of low-level nodes is performed first, and the hash strings of the referred low-level nodes are substituted into the contents of high-level nodes for obtaining the hash strings of the high-level nodes. A certain hash function like SHA-256 is called to obtain the hash strings of nodes.

Using the hash string as input, some commonly used hash functions can be called to fast map the string into an integer hash code (Estébanez, et al., 2013). In our implementation, the DJB function is used. Let $c_i$ be the ASCII value of the $i$-th character in a string, and $\text{hash}_0 = 5381$, the hash code is iteratively calculated as:
$$\text{hash}_{i+1} = \text{hash}_i * 33 + c_i . \tag{1}$$

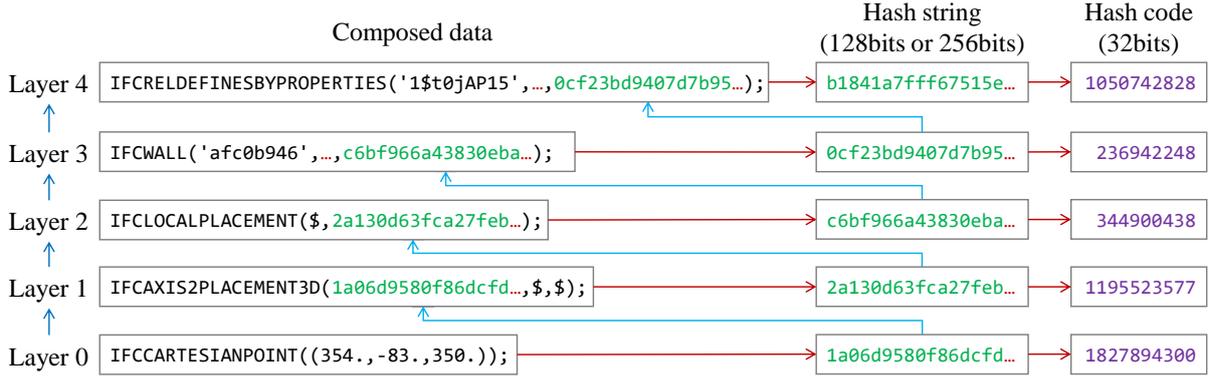

Figure 1: Example of layered hash string calculation.

### 3.2 Hash Collision Resolution in Prefix Spaces

When assigning the objects to an integer-indexed hash table with a certain capacity, collision resolution must be considered. Once a location is occupied by a previous object, the object needs to jump to the next location until a free location is found. In our implementation, the simple linear probing method is used, i.e. the object always goes to the succeeding location.

Due to the hash collision resolution, the order of scanning the input objects will affect the result of the assigned locations. This becomes a major challenge when trying to assign the hash codes in parallel, since the order of the input objects is likely to be shuffled in different runs. Aiming at the challenges of stable hash collision resolution in parallel computing, this algorithm applies a mechanism of pre-allocation of prefix spaces, so that hash collision can be stably resolved within each prefix space in single thread, and the calculation of all prefix spaces can be performed in parallel.

A prefix space is a node container with a constant capacity value $V$, and each container has a unique integer prefix value $p$. Inside a prefix space, each node is assigned with a unique integer suffix $s$, which is less than $V$. The whole hash code of a node is calculated as $pV + s$, which ensures the global uniqueness of the hash code in the graph. When using 32-bit hash code, the maximum number of prefix spaces is bound by $U = \text{floor}(2^{32}/V)$, which is the maximum allowed value for $p$.

For each type of node, several prefix spaces are allocated according to the total number of nodes of this type. If the number of nodes of a certain type is $n$, the number of allocated prefix spaces $m$ should be $m \geq \text{ceiling}(kn/V)$, in which $k$ is a constant spare rate greater than 1.0. The spare rate helps to reduce the probability that the nodes dispatched to the same prefix space will exceed the capacity of the prefix space, and also helps to reduce the workload in hash collision resolution. In our implementation, the spare rate is set to 2.0. A detailed discussion of the spare rate is provided in section 5.1.

### 3.3 The Steps of Parallel IFC Normalization

The proposed algorithm is with the following steps.

**(1) Node layering and parallel hash string calculation.** According to the method introduced in section 3.1, the IFC nodes are layered to perform the hash calculation. Considering that each node only refers to the nodes in lower layers, and the nodes in the same layer do not refer to each other, the nodes in the same layer can be processed in parallel. With the calculated hash strings, redundant nodes in the graph can be merged, so the total number of nodes in the output normalized file may be less than the original file.

**(2) Initializing the prefix spaces.** The number of prefix spaces $m$ is decided for each node type, and the summation of all $m$ values should not exceed the upper bound $U$. To decide the prefix codes for the prefix spaces, all the prefix spaces are named with a string in the format "TypeName_SerialNumber", for example, "IfcWall_0", "IfcWall_1", "IfcWall_2", etc. The names are sorted, and an integer hash code can be calculated for each name with the hash function in Eq. (1). The prefix code is calculated with the hash code modulo $U$, and linear probing is used to find a free prefix code if the calculated prefix code is occupied. The algorithm ensures that the same node type will always be assigned to the same prefix spaces in multiple runs, and the allocated prefixes are stable within a reasonable range of modification of the input IFC file. A detailed discussion of the stability of prefix allocation is provided in section 5.2.

**(3) Dispatching nodes to prefix spaces.** With the integer hash function in Eq. (1), the hash codes for all the nodes can be calculated according to the hash strings. With the hash code modulo the corresponding $m$ value of the node type, the nodes are dispatched into the prefix spaces. This step ensures that the same node will always be assigned to the same prefix space in multiple runs, and possible hash collision will only occur within the same prefix space.

**(4) Suffix assignment in each prefix space.** The suffix assignment calculation of all prefix spaces can be performed in parallel, and inside each prefix space, the nodes are sorted for stable hash collision resolution. The nodes are sorted according to the integer hash codes, and if some nodes are with the same hash code, they are compared according to the full hash strings. After sorting, the suffix is obtained with the hash code modulo the capacity $V$. If the suffix is already occupied by a previous node, the integer hash collision is resolved by linear probing.

**(5) Reference update and save.** The new row ID is decided by linking the prefix $p$ and suffix $s$ as $pV + s$, and then the references of the node in the whole graph are updated. In exporting the result IFC data, all rows are sorted according to the new row IDs. Since the row ID is assigned according to the content of the subgraph of each node, if the content of the subgraph is not changed, the data rows are likely to present in the same locations in the result IFC file, so that the text-based comparison and version control tools can identify the changes correctly.

### 3.4 Special Strategies for Better Handling IFC Schema

The proposed normalization algorithm in this paper is supposed to be applicable not only to IFC but also to various data schemas in the DAG structure. However, the following special strategies for IFC schema are implemented to obtain better normalized IFC files for getting a more reliable version comparison.

**(1) Dealing with unstable GUIDs.** The current IFC exporting tools can usually keep unchanged GUIDs for IfcElement nodes in multiple runs, but for other types of nodes (such as IfcSpatialElement, IfcObjectType, IfcProperty and IfcRelationship), the GUIDs may be unstable. To address this issue, in our implementation, the GUIDs for such types of nodes are allowed to be re-assigned by encoding the content-based hash strings into GUIDs.

**(2) Dealing with unstable "IfcOwnerHistory".** Many current IFC exporting tools write different timestamps into IfcOwnerHistory nodes in multiple runs, which causes hash string change of all referring IfcRoot nodes, hence the result row IDs are also changed. In our implementation, two optional strategies can be chosen to address this issue. One option is to ignore the IfcOwnerHistory references on calculating the hash strings, and the references are to be added back finally on saving the normalized IFC file. Hence every referring node has unchanged row ID, but only in-place reference change to the IfcOwnerHistory. Another option is to drop all references to IfcOwnerHistory, which may result in even fewer changed rows.

**(3) Dealing with important inverse edges.** IFC schema is defined as a DAG structure. Some information is accessed via the inverse edges, but such information is not included in the hash string calculation. In some special cases, the ignorance of some important inverse edges may cause unwanted merge of nodes, which results in confused data in the normalized IFC file. An example case is shown in Figure 2. The IfcTriangulatedFaceSet nodes #101 and #201 are with the same hash string, which means that they are in the same shape. But the two nodes are assigned with different colors via "StyledByItem" inverse edges by nodes #103 and #203, respectively. Since the hash strings of #101 and #201 are the same, they will be identified as redundant nodes and merged into one node, which results in confused color assignment in the normalized file. To address this issue, in our implementation, the "StyledByItem" is marked as a type of "important inverse edge", and such edges are used to update the hash strings at the end of hash calculation, so that the nodes #101 and #201 will keep separated in the output.

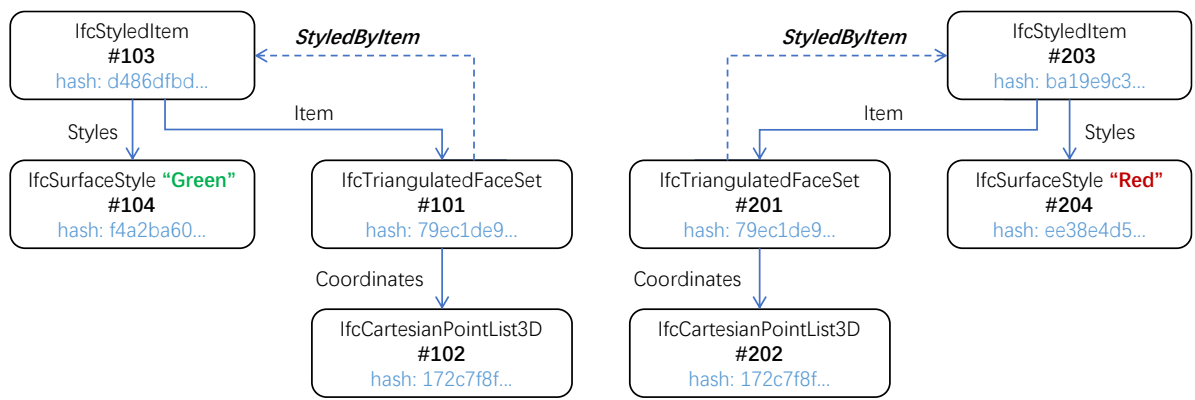

Figure 2: An example case of the important inverse edge "StyledByItem".

## 4. Experiments

In this section, two experiments are performed to show the feasibility and efficiency of the proposed normalization algorithm in IFC version control. The experiments are performed on a PC with 3.19GHz 16-core CPU and 128GB memory.

**(1) Version control of normalized IFC file versions on Git.** A series of IFC file versions are normalized using the proposed algorithm, and committed to a Git repository. The changes in storage size are recorded, and the time consumptions in normalization are compared. The original model is an Autodesk Revit file (.rvt) with 115MB. The tested IFC file versions are as follows.

- V0: exporting the IFC file from the original .rvt file.

- V1: exporting the IFC file again from the original .rvt file.

- V2: removing 5 objects and adding 5 new objects on the .rvt file, then exporting to IFC file.

- V3: changing 50 property values in-place on the exported IFC file in V2.

Table 1 shows the comparison of Git repository changes in committing the original and normalized IFC file versions, respectively. The algorithm is set as "drop references to IfcOwnerHistory", and all the ".git" sizes are recorded after executing the "git gc" command. The results show that the normalization algorithm helps Git identify the true changes, and also reduces the repository size in incremental storage. From V0 to V1, two timestamp changes in header and in "IfcOwnerHistory" are correctly identified. From V1 to V2, the 5 removed and 5 added "IfcElement"s with related representations and properties are identified as row changes by Git. From V2 to V3, in the case of in-place change in IFC, the row changes increase after normalization since the nodes referring to the changed nodes are also involved.

Table 1: Comparison of Git repository changes for original and normalized IFC file versions.

| Versions | Original IFC file | | | | Normalized IFC file | | | |
|---|---|---|---|---|---|---|---|---|
| | IFC file size (KB) | Rows | Row changes | ".git" size (KB) | IFC file size (KB) | Rows | Row changes | ".git" size (KB) |
| V0 | 28,861 | 297,735 | / | 4,233 | 14,131 | 110,108 | / | 2,712 |
| V1 | 28,861 | 297,735 | +18,659 -18,659 | 4,702 | 14,131 | 110,108 | +2 -2 | 2,713 |
| V2 | 28,837 | 297,447 | +295,339 -295,627 | 8,438 | 14,119 | 109,985 | +166 -289 | 2,720 |
| V3 | 28,837 | 297,447 | +50 -50 | 12,174 | 14,119 | 109,985 | +136 -136 | 2,733 |

**(2) Efficiency of the algorithm with different speeding-up strategies.** The normalization algorithm is tested on several larger IFC files to show the efficiency of the algorithm. Table 2 lists the time usage of normalizing 4 IFC files in different sizes. The time for loading the files and writing the normalized files to disk is excluded. The results show that the algorithm can finish normalizing medium-sized IFC files in several seconds, and finish normalizing large IFC files with around 1GB in less than 1 minute. When running in multi-threads, the algorithm can have around 30% speed-up compared with single-thread and can obtain stable results.

Table 2: Time usage of normalizing 4 IFC files in different sizes.

| File ID | Original IFC file size (KB) | Normalized IFC file size (KB) | Time usage (s) | |
|---|---|---|---|---|
| | | | multi-threads | single-thread |
| 1 | 53,483 | 22,978 | 1.40 | 2.25 |
| 2 | 197,851 | 67,634 | 4.90 | 7.85 |
| 3 | 861,193 | 378,190 | 33.09 | 46.88 |
| 4 | 1,705,112 | 686,523 | 56.66 | 86.58 |

## 5. Discussions

### 5.1 Setting the Spare Rate

The spare rate $k$ introduced in section 3.2 is applicable for a node type when the corresponding number of prefix spaces $m > 1$. The nodes are almost randomly allocated into the prefix spaces, so there is a chance that the number of nodes allocated into a prefix space exceeds the capacity

$V$ in an extremely unbalanced allocation. On assigning $n$ nodes into $m$ prefix spaces, the number of assigned nodes in a prefix space follows the binomial distribution $B(n, 1/m)$. Since $n$ is large, the binomial distribution can be approximated with the normal distribution $N(\mu, \sigma^2)$, where $\mu = n/m$ and $\sigma^2 = n(m-1)/m^2$. With the approximation, the probability that the number of nodes exceeds the capacity $V$ is the area to the right of $z = (V - \mu)/\sigma$ in a standard normal distribution. Let $\Phi(x)$ be the integral of the standard normal distribution curve, the area to the right of $z$ is $1 - \Phi(z)$. By substituting $mV \geq kn$ according to $m \geq \text{ceiling}(kn/V)$, the formula can be transformed to

$$z = \frac{mV - n}{\sqrt{(m-1)n}} \geq (k-1)\sqrt{n/(m-1)}. \tag{2}$$

So for a certain prefix space, the probability of exceeding the capacity is not greater than $1 - \Phi((k-1)\sqrt{n/(m-1)})$. A larger $k$ value will reduce the probability of exceeding the capacity of a prefix space, but the total number of nodes in the whole graph will be reduced to less than $2^{32}/k$. When $k = 1.0$, i.e. no spare for a prefix space, since $\Phi(0) = 1/2$, one prefix space has half the chance to exceed the capacity. In our implementation, the spare rate is set as $k = 2.0$, and the probability of exceeding the capacity is now $1 - \Phi(\sqrt{n/(m-1)})$. Considering that $n \gg m$, the probability can be reduced to very small, so $k = 2.0$ is a proper setting. As a backup operation, the nodes will be moved to the succeeding prefix space if the target prefix space is already full (which never happens in our experiments).

### 5.2 Stability of Prefix Allocation

In the proposed algorithm, each node type is assigned with $m$ prefix spaces, and the nodes are dispatched according to the hash codes modulo the $m$ value of this node type. When the IFC data is modified from a previous version, if the number of nodes changes in a reasonable range that the $m$ values for all types do not change, the allocation of prefix spaces and the dispatching of nodes will be stable. But if the $m$ value of a certain type is changed, the nodes of this type are likely to be dispatched to different prefix spaces, and the final row IDs will be changed. Sometimes the change of $m$ of a type may cause the change of a prefix code owned by another type, due to collision resolution in initializing the prefix spaces.

If $m$ is directly set as $m = \text{ceiling}(kn/V)$, in the case when the $n$ value (the number of nodes of this type) is close to integer times of $V/k$, the $m$ value will be changed once the $n$ value crosses the "stair edge". In order to reduce the frequency of $m$ changing, some techniques can be used. One simple way is to increase the capacity $V$, so that the intervals are enlarged. Another possible way is to scale $m$ up by a non-linear function, such as setting $m$ as the first value in the sequence $\{1, 2, 4, 8, 16 ...\}$ that is greater than $\text{ceiling}(kn/V)$. The change of $m$ only makes trouble in the text-based comparison. When such a case happens, the IFC file is still well-formed, and the comparison method by subtracting the sets of all hash strings is still applicable.

## 6. Conclusion and Future Work

The IFC normalization algorithm can stably reorganize the rows and references, so that the true modifications can be correctly identified by Git-like tools. The proposed method shows its potential for efficient incremental storage in CDE applications. In the future, the Git-based collaboration for IFC is to be further studied, including automatic branch merging and interactive conflict resolution in the workflow involving multiple users.

## Acknowledgment

This work was supported by the Science and Technology Research and Development Program of China State Construction Engineering Corporation "Research and Application of Key Technologies for Building Materials Quality Traceability Based on Independent and Controllable Blockchain" (CSCEC-2022-Z-5).

## References

Tao, X., Das, M., Liu, Y. and Cheng, J.C.P. (2021). Distributed common data environment using blockchain and Interplanetary File System for secure BIM-based collaborative design. Automation in Construction. Elsevier, 130, p. 103851.

Jaskula, K., Kifokeris, D., Papadonikolaki, E. and Rovas, D. (2022). Common Data Environments in construction: State-of-the-art and challenges for practical implementation. SSRN preprint (October 16, 2022). Available at: http://dx.doi.org/10.2139/ssrn.4249458.

buildingSMART (2019). OpenCDE API standards. https://github.com/buildingSMART/OpenCDE-API/, accessed February 2023.

Preidel, C., Borrmann, A., Oberender, C. and Tretheway, M. (2017). Seamless integration of common data environment access into BIM authoring applications: The BIM integration framework. In: eWork and eBusiness in Architecture, Engineering and Construction. CRC Press, pp. 119-128.

Patacas, J., Dawood, N. and Kassem, M. (2020). BIM for facilities management: A framework and a common data environment using open standards. Automation in Construction. Elsevier, 120, p. 103366.

Werbrouck, J., Pauwels, P., Beetz, J. and van Berlo, L. (2019). Towards a decentralised Common Data Environment using linked building data and the Solid ecosystem. In: 36th CIB W78 2019 Conference, pp. 113-123.

Senthilvel, M., Oraskari, J. and Beetz, J. (2020). Common Data Environments for the Information Container for linked Document Delivery. In: Proceedings of the 8th Linked Data in Architecture and Construction Workshop-LDAC, pp. 132-145.

Firmenich, B., Koch, C., Richter, T. and Beer, D.G. (2005). Versioning structured object sets using text based Version Control Systems. In: Proceedings of the 22nd CIB-W78.

Nour, M., Firmenich, B., Richter, T. and Koch, C. (2006). A versioned IFC database for multi-disciplinary synchronous cooperation. In: Proceedings of the Joint International Conference on Computing and Decision Making in Civil and Building Engineering, Montreal, Canada, pp. 636-645.

Poinet, P., Stefanescu, D. and Papadonikolaki, E. (2020). Collaborative workflows and version control through open-source and distributed common data environment. In: Proceedings of the 18th International Conference on Computing in Civil and Building Engineering: ICCCBE 2020, pp. 228-247.

Esser, S., Vilgertshofer, S. and Borrmann, A. (2022). Graph-based version control for asynchronous BIM collaboration. Advanced Engineering Informatics. Elsevier, 53, p. 101664.

Li, S., Gao, G., Wang, W., Liu, H., Zhu, S. and Gu, M. (2022). A redundancy-free IFC storage platform for multi-model scenarios based on block hash. In: Proceedings of the EG-ICE 2022 Workshop on Intelligent Computing in Engineering, pp. 74-83.

Shi, X., Liu, Y.S., Gao, G., Gu, M. and Li, H. (2018). IFCdiff: A content-based automatic comparison approach for IFC files. Automation in Construction. Elsevier, 86, pp. 53-68.

Bruno P. (2022). Native IFC whitepaper. https://github.com/brunopostle/ifcmerge/blob/main/docs/whitepaper.rst, accessed February 2023.

Estébanez, C., Saez, Y., Recio, G., and Isasi, P. (2013). Performance of the most common non-cryptographic hash functions. Software: Practice and Experience. Wiley Online Library, 44(6), pp. 681-698.